\documentclass[12pt]{article}
\usepackage{graphics}
\def\ea{{\em et al.}}
\def\ni{\noindent}
\def\ph{{\phantom{...}}}
\def\={\phantom{..} = \phantom{..}}

\def\+{\phantom{..} + \phantom{..}}
\def\>{\phantom{..} > \phantom{..}}
\def\<{\phantom{..} < \phantom{..}}
\def\-{\phantom{..} - \phantom{..}}

\def\lra{\longrightarrow}

\def\lsb{\left[\,}
\def\rsb{\,\right]}

\def\rng{random number generator}
\def\inter{intervention}
\def\ie{intervention-efficacy}

\def\pop{population}
\def\pops{populations}

\def\rv{random variable}
\def\rvs{random variables}

\def\bp{branching-process}
\def\bps{branching-processes}

\def\prob{probability}
\def\probs{probabilities}

\def\sto{stochastic}
\def\epv{extra-Poisson variation}

\def\ro{R_0}
\def\vo{V_0}
\def\rl{R_l}
\def\rh{R_h}
\def\ph{p_h}
\def\pl{p_l}
\def\pe{p_e}
\def\pei{p_{e,i}}

\def\no{\nonumber}
\def\be{\begin{equation}}
\def\ee{\end{equation}}

\def\bar{\begin{eqnarray}}
\def\ear{\end{eqnarray}}
\def\bad{\begin{equation} \begin{array}}
\def\ead{\end{array} \end{equation}}
\def\bmat{\left( \begin{array}}
\def\emat{\end{array} \right)}

\title{\bf \scalebox{1.5}{Stopping the SuperSpreader}\\ 
\scalebox{1.5}{Epidemic, Part II:}\\[1in] 
\scalebox{1.5}{MERS Goes Pandemic}\\[2in]}

\author{W. David Wick$^1$\footnote{
Email:wdavid.wick@gmail.com}} 

\begin{document}
\maketitle
\pagebreak

\section*{Abstract}
In a paper of August 2013, I discussed 
the so-called SuperSpreader (SS) epidemic model and emphasized that it has dynamics differing greatly from the more-familiar uniform (or Poisson) textbook model.
In that paper, SARS in 2003 was the representative SS instance and it was suggested that MERS may be another. 
In April of 2014, MERS incident cases showed a spectacular spike (going from a handful
in the previous April to more than 260 in that month of 2014) reminiscent of a figure I published nine months previously. 
Here I refit the two-level and several variant SS models to incident data from January 1, 2013--April 30, 2014 and conclude that MERS will cause a pandemic (all other factors remaining the same).
In addition I discuss a number of model-realism and fitting-methodology issues relevant to analysing SS epidemics. 

\pagebreak

\section*{Text}

As in the previous paper, \cite{augustpaper}, 
I avoid the standard journal format (Introduction/Methods/Results/Discussion) for ease of exposition.
There are no mathematical prerequisites for reading this paper.

\begin{figure}
\rotatebox{0}{\resizebox{5in}{5in}{\includegraphics{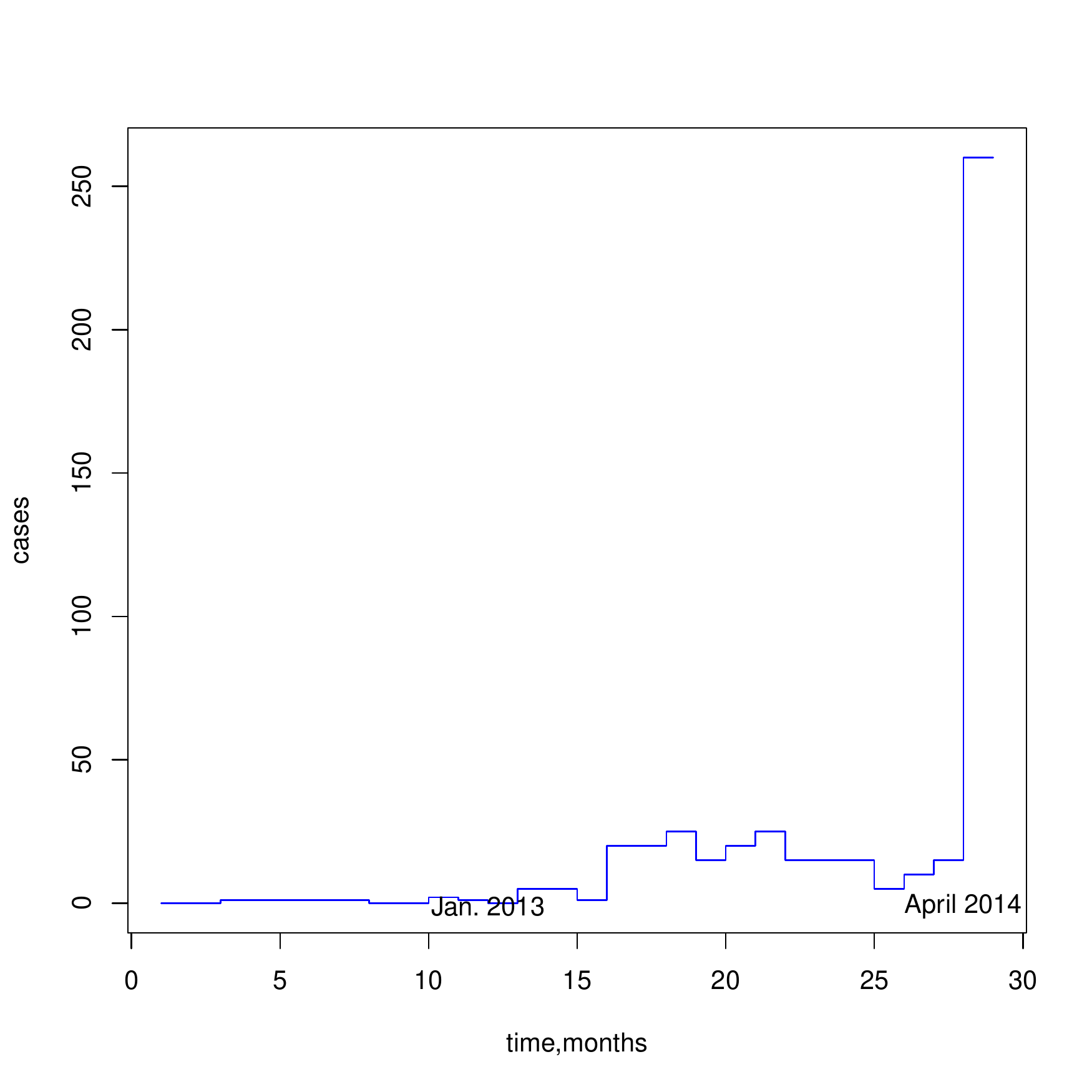}}}
\caption{MERS incident case data (per month), adapted from a ECDC figure.\label{ECDCfig}}
\end{figure}

To begin, look at Figure \ref{ECDCfig}, adapted from the European Center for Disease Control (ECDC) website on May 5, 2014, showing incident cases of the 
Middle East Respiratory Syndrome (MERS, also abbreviated MERS-CoV because it is caused by a novel coronavirus).
The ECDC explained that the gigantic increase in cases, more than 10-fold, in April of 2014 might be explained by:

\begin{itemize}

  \item  More sensitive case detection through more accurate case finding and contact tracing \ldots 
  \item Increased zoonotic transmission \ldots
  \item Breakdown in infection control measures \ldots in the local health care setting \ldots
  \item Change in the virus resulting in more effective human-to-human transmission \ldots
  \item False positive lab results.
\end{itemize}

Concerning the fourth listed, a nearly simultaneous report by virologists sequencing the virus found no suspicious mutations.
While all of these explanations are possible, the ECDC left out one: chance. 

\begin{figure}
\rotatebox{0}{\resizebox{5in}{5in}{\includegraphics{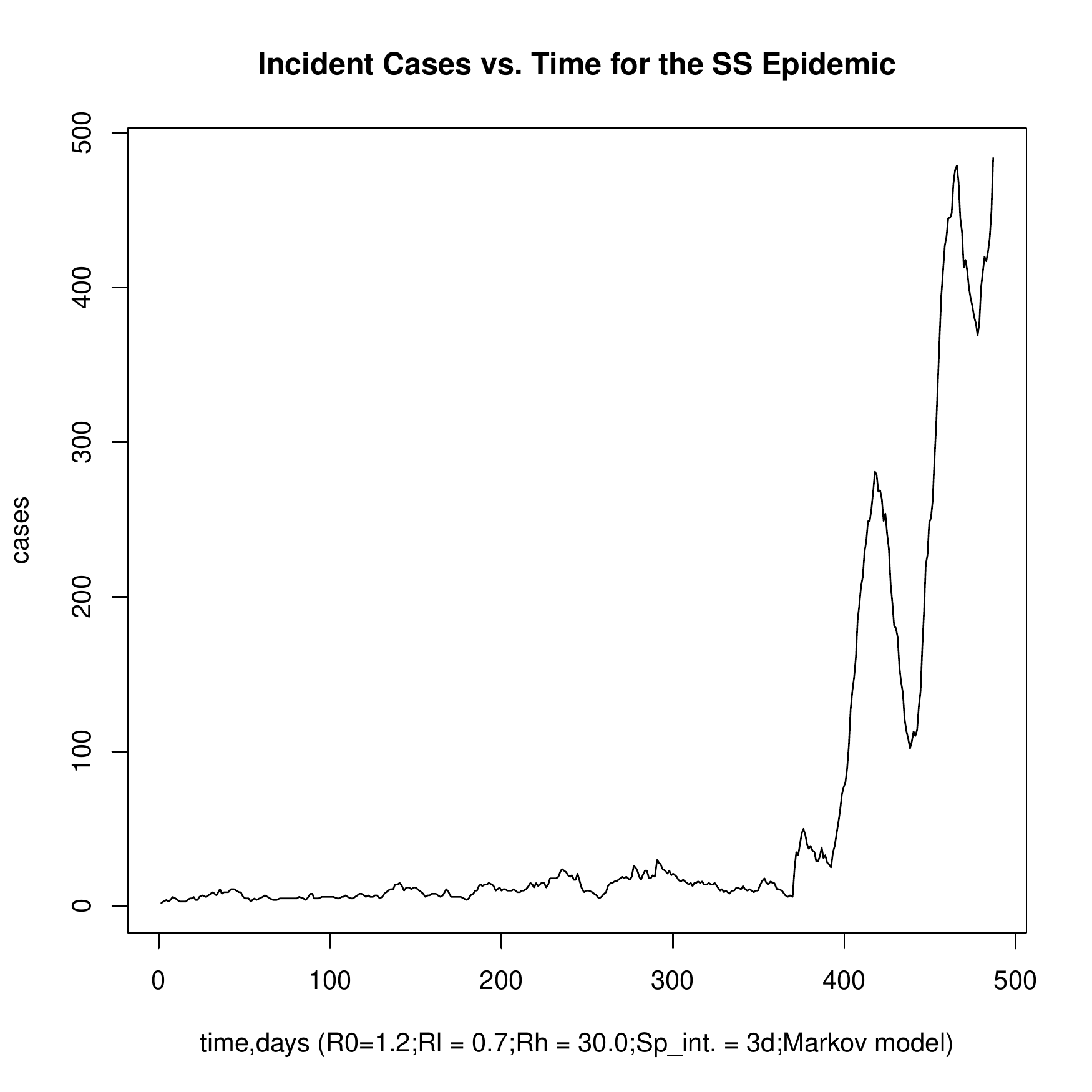}}}
\caption{An SSe with large $\vo$ but lower $\ro$, and sporadic cases.\label{Figcase2}}
\end{figure}

For instance, there is Figure \ref{Figcase2}, which I published 9 months before, showing simulated cases (in hospital) for an SS epidemic with $\ro$ slightly higher than one. 
Why the strange delay and then the sudden spike? As explained in that paper, an SS epidemic is stochastic and may have a long ``kindling time'' before it takes off. 
I emphasize that no model parameter was allowed to change with time during this simulation; the surprising pattern is due solely to chance events during the epidemic.

\begin{figure}
\rotatebox{0}{\resizebox{5in}{5in}{\includegraphics{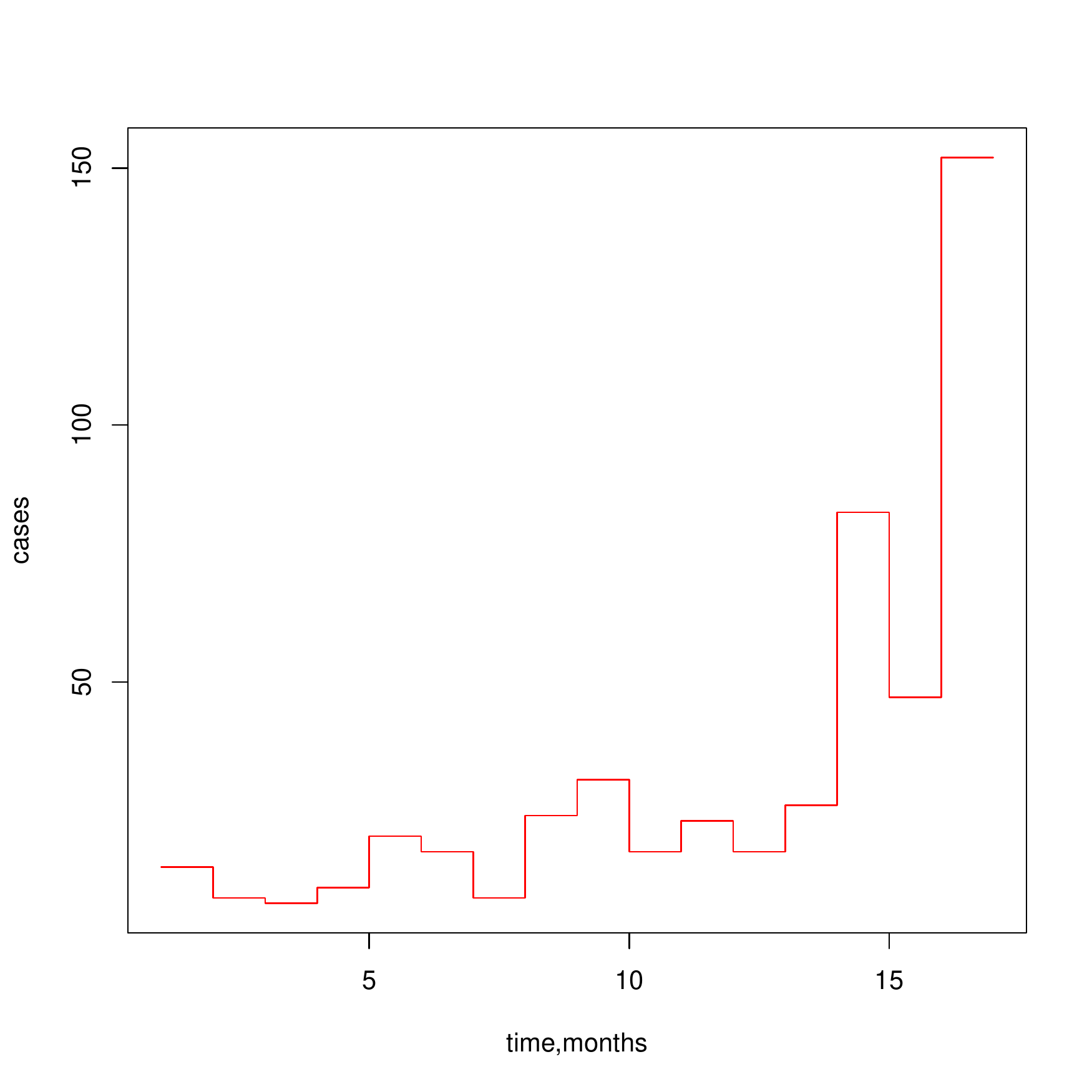}}}
\caption{Simulated MERS incident cases (per month), from the GOFS run of Model A.\label{simfig1}}
\end{figure}

\begin{figure}
\rotatebox{0}{\resizebox{5in}{5in}{\includegraphics{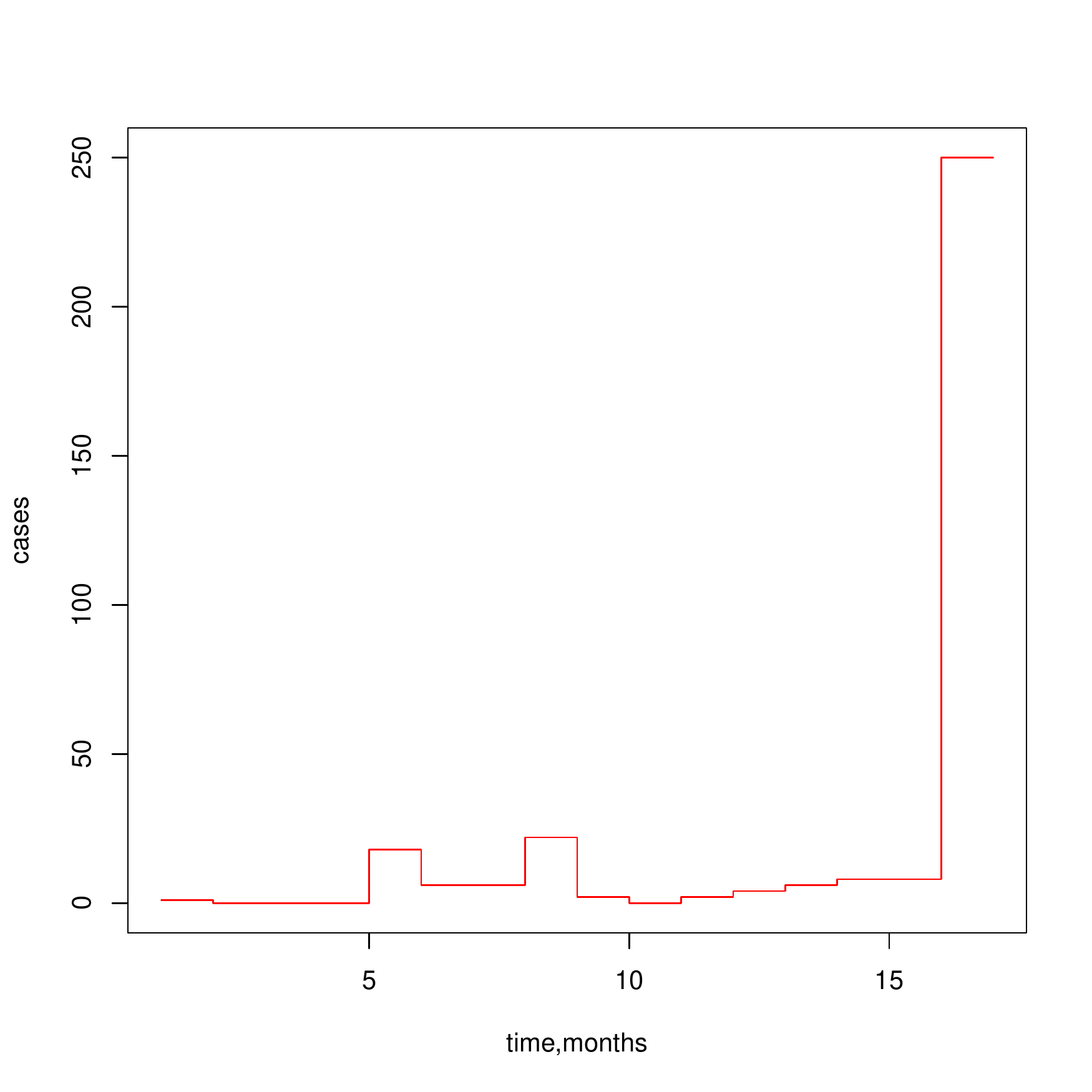}}}
\caption{Simulated MERS incident cases (per month), from the GOFS run of Model B.\label{simfig2}}
\end{figure}

\begin{figure}
\rotatebox{0}{\resizebox{5in}{5in}{\includegraphics{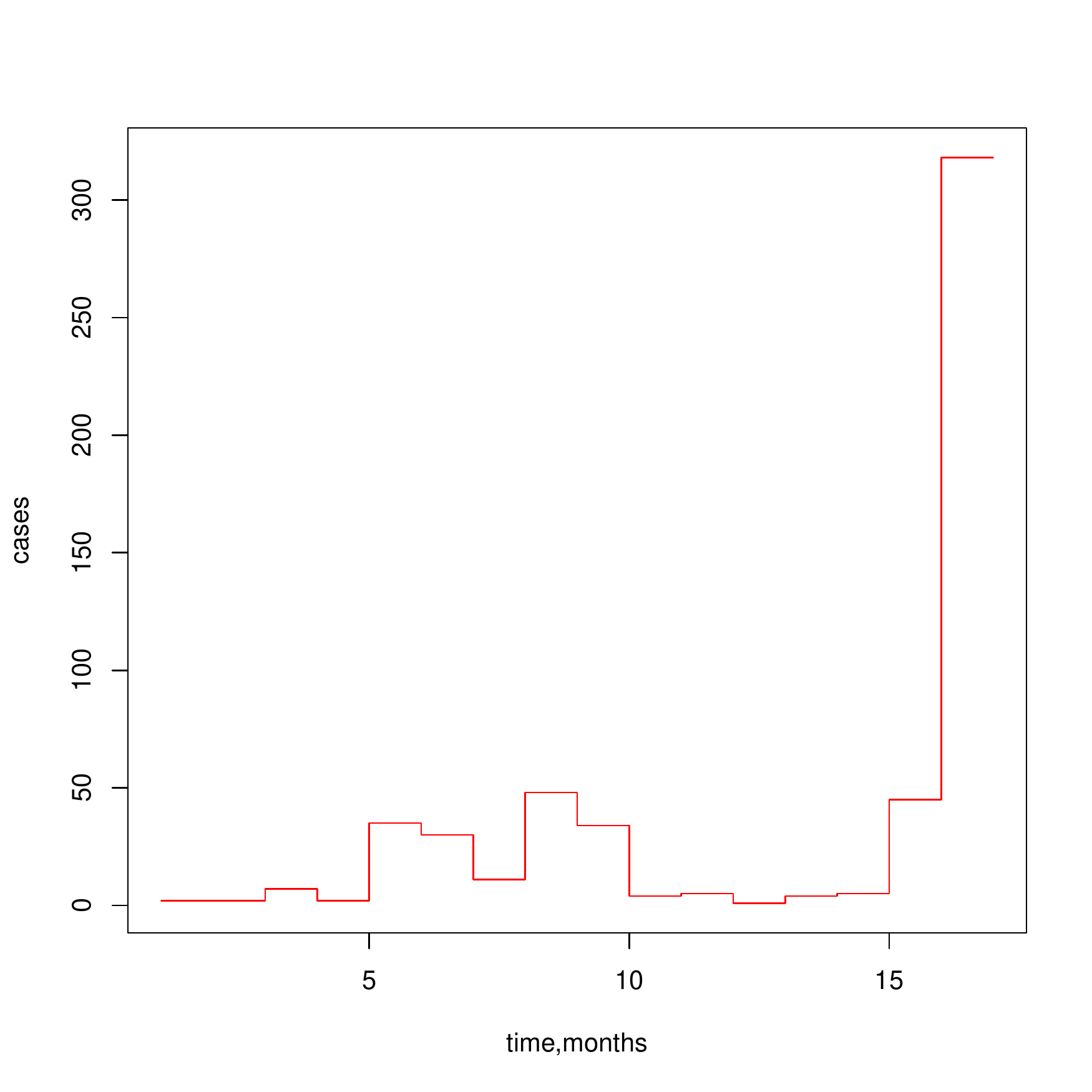}}}
\caption{Simulated MERS incident cases (per month), from the GOFS run of Model C.\label{simfig3}}
\end{figure}

Now consider Table 1, reporting model fits for three classes of SS epidemic models to the ECDC data during Jan. 2013--April 2014. 
Look first just at the column reporting $\ro$---the ``basic reproductive number'' of the infection, or average number of secondary cases caused by one infected case during the time while
remaining infectious. I will explain the models and other entries below. Next examine Figures 
\ref{simfig1}--\ref{simfig3} for the simulated epidemics, showing cases per month as in Fig. \ref{ECDCfig} but for 2013--April 2014 only, for a certain ``run'' 
of the model called the GOFS, explained below.

A model with $\ro > 1$ displays, eventually, exponential growth, so the meaning is: pandemic.

\centerline{}
\centerline{}
\centerline{}
\centerline{Table 1. Results of fitting three SS models }
\centerline{to the ECDC MERS incidence data, 2013--April 2014}
\begin{center}
\begin{tabular}{|c|c|c|c|c|c|c|}\hline
Model & $\ro$ & $\rl$ & $\rh$ & sporadic time$^*$ & alpha & GOFS$^{**}$ \\ \hline
A & 1.0 & 0.49  & 58.0  & 3.1 & - & 137.5  \\ \hline
B & 1.44  & 0.30  & 55.0 & -  & - & 45.7  \\ \hline
C & 1.44 & 0.30  &  - & -  & 1.31 & 78.7  \\ \hline
\end{tabular}
\centerline{}
$^{*}$ average time between sporadic cases, days\\
$^{**}$ square-root of sum-of-squares of simulated\\ minus observed cases, per month
\end{center}
\centerline{}
\centerline{}
\centerline{}
   
\begin{figure}
\rotatebox{0}{\resizebox{5in}{5in}{\includegraphics{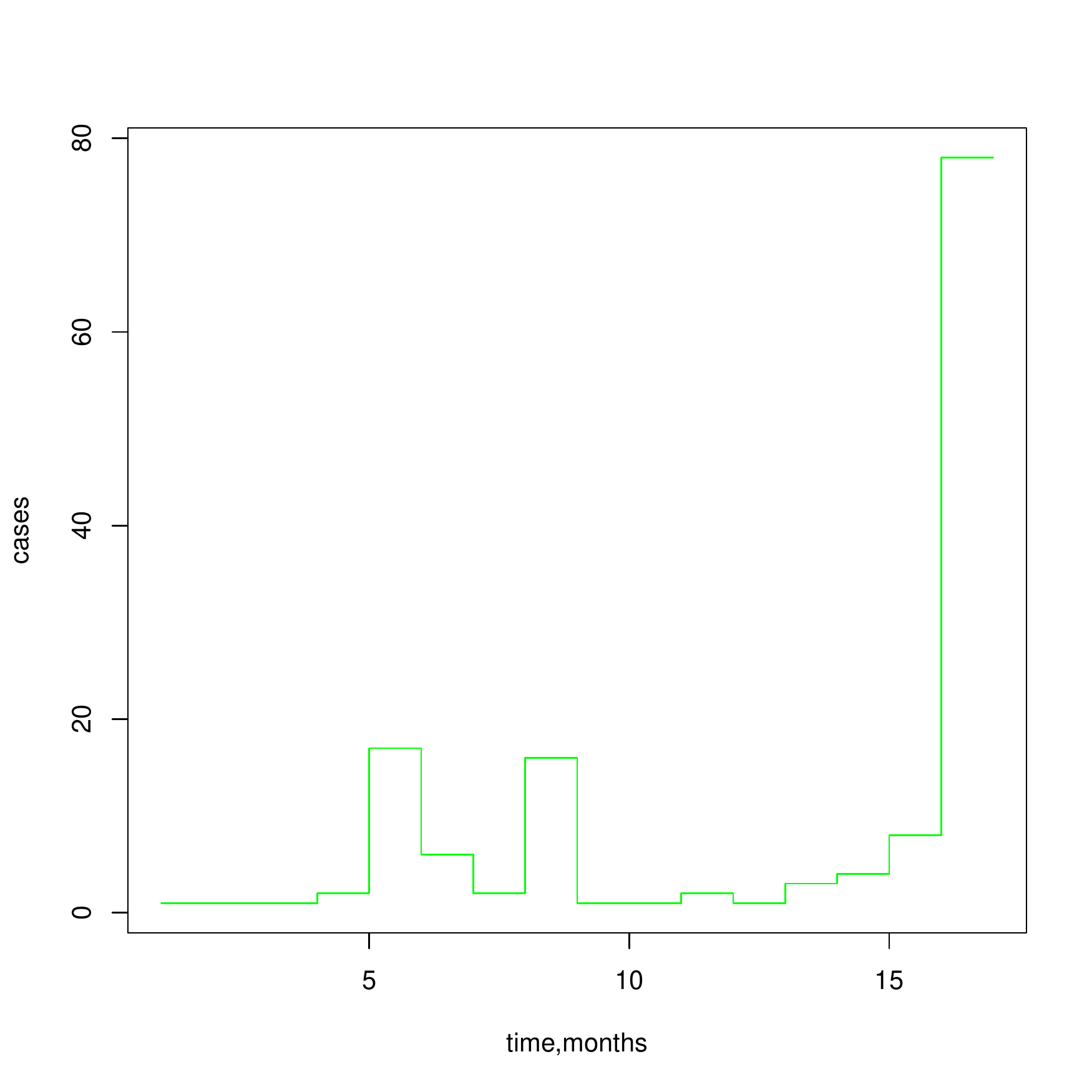}}}
\caption{My stab at a time-dependent sporadic incidence rate, based on annotations in FluTrackers's case list.\label{primaryfig}}
\end{figure}

Now to describe the three models in the fitting exercise. Model A is the two-level model of \cite{augustpaper}, 
with a ``low spreader'' (secondary cases with mean $\rl$) and ``high spreader'' (ditto but mean $\rh$) appearing with probabilities arranged so that the overall mean of reproduction is $\ro$.
For Model B I tried to use what information is currently available that can distinguish primary (sporadic, zoonotic exposure) 
cases from cases due to human contacts, in order to allow for an (unexplained; seasonal or other) increase in the sporadic case rate. 
So I accessed the FluTrackers website, MERS case (hospitalization) list page, 
on May 8, 2014, and dropped all the entries that contained any of the following annotations:
foreign (meaning not Middle East); health-care worker (HCW), contact of confirmed case, multiple co-morbidities (reasoning that such a patient was more likely to have contracted MERS
in hospital than at a camel farm) or recently in hospital (within 10 days of hospitalization for MERS). I then derived a sporadic-incidence rate, cases per month, see Fig. \ref{primaryfig}, 
and modified the
model by adding it as a time-dependent covariate. Finally, for Model C I adopted a variable $\rh$ with a power-law distribution, described below, and kept the time-dependent sporadic rates. 
The reason for the coincidences in Table 1 (e.g., 1.44 and 0.30 appearing twice) just reflects a limit on the number of models
considered for practical reasons; 
I divided a range (e.g., $\ro$ in $[0.6,2.0]$; $\rl$ between 0.3 and either $\ro$, if $\ro < 1$, or one; $\rh$ in $[10,60]$; sporadic time in $[1,6]$; alpha in $[1.1,10]$) into 8-10
parts and tested all the parameter combinations. 
Fixed parameters were: mean noninfectious period, 5.36 days; infectious period, 3.83 days (taken from fitting a SARS model to 2003 data, \cite{book08})
with standard deviations (log-normal distributed) of twice the means. 

Why choose a variable $\rh$ for Model C? I thought my two-level model a bit simplistic.
Remember from the SARS story the Metropole Hotel, the Prince of Wales Hospital, the Amoy Gardens and that plane to Beijing (see, e.g., \cite{book08} Chapter 6, for references). 
Perhaps $\rh$, instead of a fixed value,
 should have some long-tailed distribution. So I programmed an alternative model with a power law:

\be
P\lsb \rh > x \rsb \= \left(\,{1\over x^{\alpha}}\,\right)\,1\left[\,x > 1\,\right].
\ee

This rather extreme choice has the pecularity that, for $1 < \alpha < 2$, $\rh$ has finite mean but infinite variance. This is not necessarily absurd (recall the 300+ secondary cases of
one infected person at the Amoy Gardens apartment complex in Hong Kong) as some finite upper bound would apply in any case. 

A note on the fitting method. I use the famous ``kNN'' (k-th nearest neighbor) method from computer science, 
which dates from the 1950s; it is usually applied to pattern-recognition problems in which some
samples (of, say, pictures of cats, horses, or people) are available and the goal is to classify a given sample (picture) into one of the categories. I reversed the idea by 
taking the known samples to be simulations from my SS epidemic model with various parameter combinations, with the sample to be classified coming from nature. So the interpretation 
becomes: which simulation sample most resembles the natural sample? The goodness-of-fit statistic (GOFS) 
I used is the sum-of-squared-differences (or its square-root) in monthly incident cases (more on that choice below) of
the k-th nearest neighbor simulation sample cases, minus the actual cases, 
with $k = N^{4/9}$ (the optimal exponent as proven in the book \cite{book08}) and usually $N = 500$ simulation samples. 
This methodology is suitable for any kind of stochastic model and observed sample, as it makes no assumptions whatsoever about distributions (e.g., no assumptions that anything is normal)
and reduces to maximum likelihood if you can simulate infinitely fast (i.e., as $N \to \infty$). So there is no loss of efficiency if your computer is speedy.

That being said, the method offers no advice on what to use for the GOFS beyond the choice of $k = N^{4/9}$. For instance, sum-of-squares of case-incidence data
 may not be the most-informative choice given the main interest. 
For the question: has MERS gone pandemic?, i.e., is $\ro >1$?, obviously an epidemiological data set identifying secondary cases, tertiary cases, etc., would be preferable, 
with the GOFS comparing
frequencies of epidemiologically-linked clusters of each size. 
In fact I used cluster data from the Lancet paper of 2013, \cite{lancet}, in my 2013 paper (but a month later used incidence data as well, with the same conclusion: not yet pandemic).
At this time I have only incidence and some informal descriptions of case status available, but I used all the data known to me in the analysis reported here.

Two other issues are topics for further research. I ignored the handful of cases from 2012, reasoning that case ascertainment early in an epidemic caused by a novel organism was poor; 
the true starting time is unknown; and that the model would try to fit all those
zeros from 2012. The last raises a curious question about prediction. Fitting the SS model to the first year-and-a-half of data led to the conclusion of \cite{augustpaper} 
that $\ro <1 $ then, but I now find that
the data was consistent with $\ro$ greater than one. Can probability statements, of the confidence-interval variety (``We are now 90 percent confident that a pandemic is inevitable \ldots''),
be derived for an SS epidemic, and if so at what time point? Can similar confidence intervals be derived for epidemic parameters? Consistent with the ``everything through simulation''
philosophy for complex models with unknown statistical properties advertised in \cite{book08}, 
this can be studied, but at present this author's PC required an overnight run for a single analysis.
More, or faster, computers are needed for the task.

On May 14, 2014, the World Health Organization (WHO) IHR Emergency Committee meeting concerning MERS-CoV declined to declare an international health emergency, asserting that there is
``no evidence of sustained human-to-human transmission,'' seemingly putting the blame on camels. (They did state that  ``the seriousness of the situation had increased.'') 
In my view, although evidence based on modeling is never as convincing as evidence based on shoe-leather epidemiology and hard facts, the Committee overlooked the stochastic nature of
an SS epidemic, which can mean that the lag of more than a year followed by a dramatic spike is consistent with a pandemic, as opposed to some unexplained event (such as a wave of camel births or
a breakdown of hospital infection-control procedures). Perhaps the risk is in the eye of the beholder. Which is the more plausible explanation for the 260+ case spike in April 2014:
chance in an SS model, or some unusual zoonotic or nosocomial happening? If the latter, we are betting on a one-off event that will not be repeated. 
If the former, we must doubt the modeling to accept
the Committee's statement;
but which assumption am I supposed to reject?

I conclude, on the basis of SS epidemic modeling and fitting, 
that MERS threatens to go beyond, or is already beyond, the pandemic threshold. 
If so, what happens next? A dip in cases in May, or even a longer period of relative quietude stretching into the summer, 
should not be taken as grounds for complacency (look again at Fig. \ref{Figcase2}).
Paper \cite{augustpaper} contains some results about stopping the epidemic through secondary case isolation.
(No drugs or vaccines exist at present for human coronaviruses.)
I can summarize by saying that the SS epidemic is actually easier to control than a uniform epidemic 
(as, e.g., influenza) because it offers a potential target for efficacious intervention,
namely if the superspreader is a consequence of context (patient in an ICU, or symptomatic case on an airplane, as just occurred on a flight to America originating in Saudi Arabia) 
as opposed to a biological factor beyond our control. 
That is, up to a point---for I also showed that 
efficacy diminshes with increasing
$\vo$, the variance of the basic reproduction number (whose mean is $\ro$). 
Nevertheless (and I admit considering a case here with $\vo = \infty$) I regard it likely that the pandemic will, like SARS
in 2003, be rapidly controlled by public health agencies acting in concert with political authorities, once the emergency is declared.

\section*{Appendix 1: some model details}

The incorporation of variable sporadic-case rates required some changes in the implementation. 
The basic simulation routine is agent-based, meaning it stores all times-to-events for all ``agents''
(cases), finds the next one, updates the time, performs any changes, and stores new times-to-events as required. 
What about reaching the end of the month? That becomes one of the stored
times-to-events; if that is the next leap forward, the time-to-next-sporadic-case is replaced by another exponential random variable 
using the rate for that month. But what if the rate hasn't changed; 
do we make a mistake? No. Recall the so-called ``bus-waiting paradox,'' which isn't really a paradox but a property of the Poisson process (which is reasonable for sporadic events) in which 
the time to the next occurrence always has an exponential distribution. The point is that the Poisson process is memoryless, so the fact that something (a new sporadic case)
didn't happen is forgettable.

The power law is easily introduced by the trick:

\be
\rh = U^{-1/\alpha},
\ee

\ni where $U$ is a uniform $[0,1]$ random variable supplied by the RNG. 
Then an infection intensity is chosen as: $[\hbox{infectious period}]/\rh$, and used to create a secondary case after an exponential
waiting time:  ($- \log(U)/\hbox{intensity}$), as usual. The fact that $\rh$ has infinite variance if $1 < \alpha < 2$ causes no trouble.  
Note that the search region in Model C is restricted by

\be
\alpha < {\ro\over \ro -1},
\ee

\ni which is necessary to assure that $\rh = \alpha/(\alpha - 1) > \ro$. 

C code, which should run on anything, is available from the author.


\begin{thebibliography}{10}

\bibitem{augustpaper}
Wick, WD. Stopping the SuperSpreader Epidemic; the lessons from SARS (with, perhaps, applications to MERS). arXiv:1308.6534v1 [q-bio.PE] 29 Aug 2013. 

\bibitem{lancet}
Breban, R, Riou, J, Fontanet, J. Interhuman transmissibility of Middle East respiratory syndrome coronavirus: estimation of pandemic risk.  
Lancet, published online July 5, 2013.

\bibitem{ploscb}
Blumberg, S and Lloyd-Smith, JO. Inference of $\ro$ and transmission heterogeneity from the size distribution of stuttering chains.
PLoS Comput Biol 2013: 9, e1002993. 

\bibitem{book08} 
Wick, WD. {\em Fitting Non-Linear, Stochastic Models to Data in Biology and Medicine}. Available on Amazon.com, 2012.


\end{thebibliography}
\end{document}